\documentclass[letterpaper]{article} 
\usepackage{aaai25}  
\usepackage{times}  
\usepackage{helvet}  
\usepackage{courier}  
\usepackage[hyphens]{url}  
\usepackage{graphicx} 
\usepackage{natbib}  
\usepackage{caption} 
\usepackage{algorithm}
\usepackage{algorithmic}
\usepackage{multirow}
\usepackage{makecell}
\usepackage{utfsym}
\usepackage{amsmath}
\usepackage{amsfonts}
\usepackage{subfigure}

\usepackage{newfloat}
\usepackage{listings}
\DeclareCaptionStyle{ruled}{labelfont=normalfont,labelsep=colon,strut=off} 
\floatstyle{ruled}
\newfloat{listing}{tb}{lst}{}
\floatname{listing}{Listing}
\title{SongEditor: Adapting Zero-Shot Song Generation Language Model as a Multi-Task Editor}
\author {
    Chenyu Yang\textsuperscript{\rm 1},
    Shuai Wang\textsuperscript{\rm 1,\rm 3},
    Hangting Chen\coauthor \textsuperscript{\rm 2 },
    Jianwei Yu\coauthor \textsuperscript{\rm 2},
    Wei Tan\textsuperscript{\rm 2},
    Rongzhi Gu\textsuperscript{\rm 2},
    Yaoxun Xu\textsuperscript{\rm 4},
    Yizhi Zhou\textsuperscript{\rm 6},
    Haina Zhu\textsuperscript{\rm 5},
    Haizhou Li\textsuperscript{\rm 1, \rm 3}
}
\affiliations {
    \textsuperscript{\rm 1}The Chinese University of Hong Kong, Shenzhen (CUHK-Shenzhen), Shenzhen, China\\
    \textsuperscript{\rm 2}Tencent AI Lab \\ \textsuperscript{\rm 3}Shenzhen Research Institute of Big Data\\
    \textsuperscript{\rm 4}Tsinghua University \\ \textsuperscript{\rm 5}X-LANCE Lab, Shanghai Jiao Tong University\\
    \textsuperscript{\rm 6}National Key Laboratory of Novel Software Technology, Nanjing University\\
    chenyuyang2@link.cuhk.edu.cn, erichtchen@tencent.com, tomasyu@foxmail.com
}

\usepackage{bibentry}

\begin{document}
\maketitle

\begin{abstract}
The emergence of novel generative modeling paradigms, particularly audio language models, has significantly advanced the field of song generation. Although state-of-the-art models are capable of synthesizing both vocals and accompaniment tracks up to several minutes long concurrently, research about partial adjustments or editing of existing songs is still underexplored, which allows for more flexible and effective production. In this paper, we present SongEditor, the first song editing paradigm that introduces the editing capabilities into language-modeling song generation approaches, facilitating both segment-wise and track-wise modifications. SongEditor offers the flexibility to adjust lyrics, vocals, and accompaniments, as well as synthesizing songs from scratch. The core components of SongEditor include a music tokenizer, an autoregressive language model, and a diffusion generator, enabling generating an entire section, masked lyrics, or even separated vocals and background music. Extensive experiments demonstrate that the proposed SongEditor achieves exceptional performance in end-to-end song editing, as evidenced by both objective and subjective metrics. 
\end{abstract}

%
\begin{links}
    \link{Demo}{https://cypress-yang.github.io/SongEditor_demo/}
\end{links}
\section{Introduction}

In recent years, applications for generating vocal music have achieved remarkable results.
Compared with speech or accompanied singing voice, a song typically contains multiple sections interspersed with music-only segments like prelude, interlude, and postlude, which requires strict consistency and contextual coherence. Therefore, an ideal song generation system must be both structure-guided and capable of handling long content. %

However, part of the generated song might be unsatisfactory in practice, but regenerating the entire sequence takes much time and computing resources, and potentially leads to further problems. Therefore, song editing is crucial as it allows for the refinement and enhancement of musical compositions. Moreover, there are instances where only a specific track, such as vocal or accompaniment, needs to be changed.  
How to rewrite the designated part of a song without destroying its contextual consistency and overall musicality, is a significant challenge for AI song composition.

 \begin{figure}[t]
  \centering
  \includegraphics[width=\linewidth]{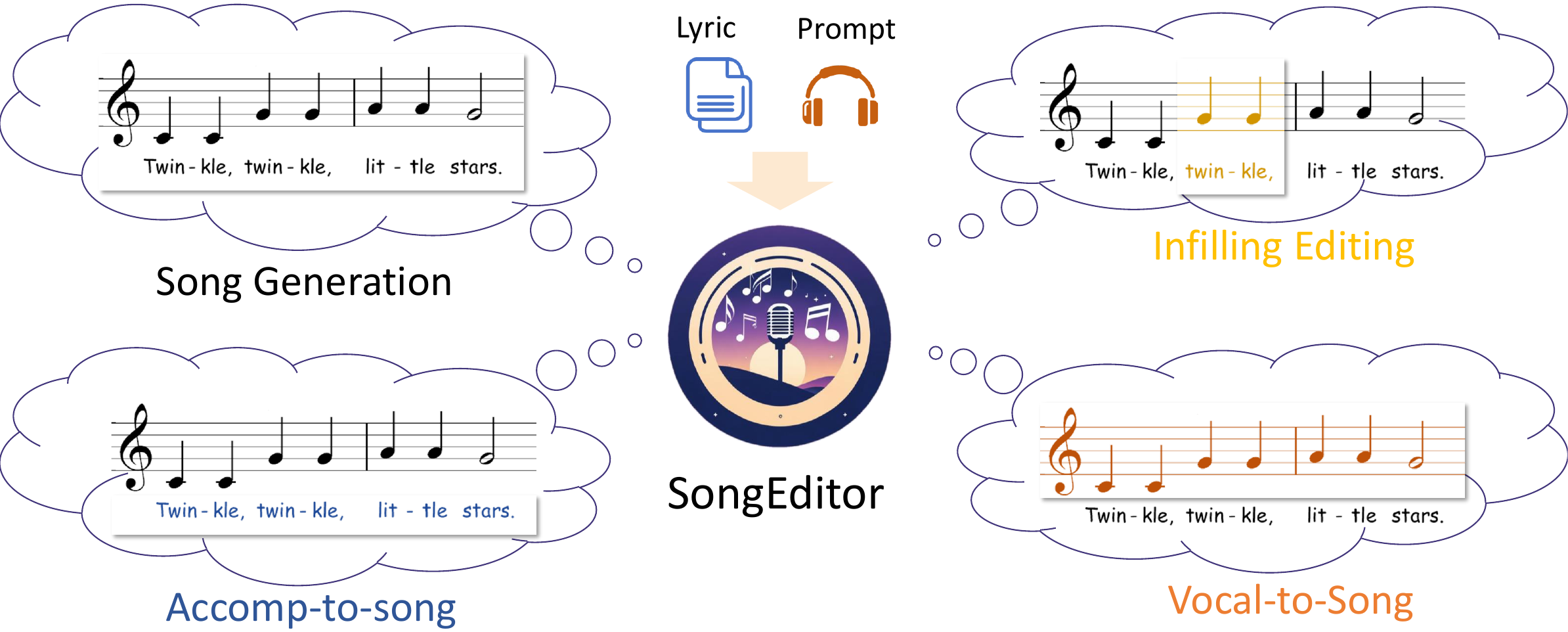}
  \caption{SongEditor supports various song generation and editing tasks, involving creating complete songs from scratch and Infilling Editing.  Accomp-to-song and vocal-to-song mean generating full songs based on partial input conditions such as accompaniments or vocals, respectively. }
  \label{fig:intro}
\end{figure}

Currently, there are relatively few studies on end-to-end song generation in the academic field. State-of-the-art song generation platforms like Suno\footnote{\url{https://suno.com}} are generally for commercial use and not open-source. Jukebox~\cite{dhariwal2020jukebox} can synthesize songs in specific styles and genres, but it lacks more refined controls such as structural adjustments or specific instructions. To date, these approaches have primarily focused on pure generation and cannot be directly applied to editing scenarios.

\citet{ding2024songcomposer} proposed SongComposer, which supports various tasks like lyric-to-melody and melody-to-lyric, potentially allowing for some degree of song editing. However,  SongComposer decomposes the song into lyrics and musical notes (MIDI), retaining no information about vocals such as timbre. Additionally, there is an unavoidable distortion between the original song and the extracted MIDI files during data preprocessing.

In this paper, we propose SongEditor, a general paradigm that enables partial editing in existing language-model-based song generation approaches.
Since few techniques of song generation model are currently available, we first implement SongLM, a fundamental zero-shot song generation framework that can generate song sections up to two minutes long based on structure-guided lyrics and a 10-second excerpt from a reference song as an acoustic prompt. Subsequently, we make modifications in the architecture design and model training process, enabling the system to perform editing based on different types of contexts.

As shown in figure~\ref{fig:intro}, in addition to generating songs from scratch, SongEditor can also regenerate specific periods and separate vocals or accompaniments. Worth noting is that the above two capabilities can be performed at the same time, allowing for the editing of even a short segment of vocals or accompaniment.
The main contributions of this paper are summarized as follows:
\begin{itemize}
\item We propose SongEditor, a language model-based generation approach that handles long-content generation guided by user-defined lyrics and structures.
\item SongEditor is among the first works designed not only to generate songs from scratch, but also to conduct sentence-level editing at the specific segment (\textit{segment-wise}) and complete vocal or accompaniment tracks when given the rest (\textit{track-wise}).
\item A context-free editing method is employed that can effectively reduce the reliance on contextual lyrics and proves to be more suitable for long-content editing. Additionally, considering the complex composition of accompaniment, a series of measures are taken to ensure the consistency of editing segments and the naturalness of transitions.
\end{itemize}

We evaluated our model using both objective metrics and human evaluations. Results show that our model can generate songs with superior musicality and excellent consistency.

\section{Related Work}

\begin{figure*}[th]
    \centering
    \subfigure[Stage 1: Train tokenizer and generator.] {\includegraphics[width=0.21\linewidth]{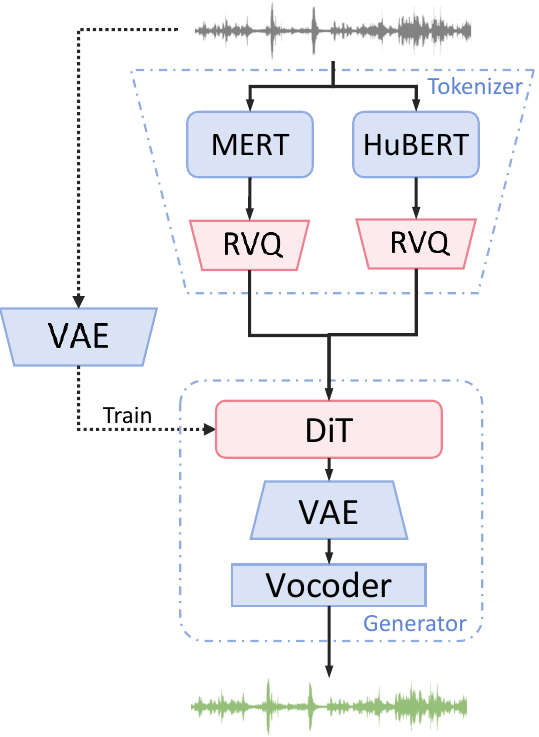}}
    \subfigure[Stage 2: Train language model.]{\includegraphics[width=0.54\linewidth]{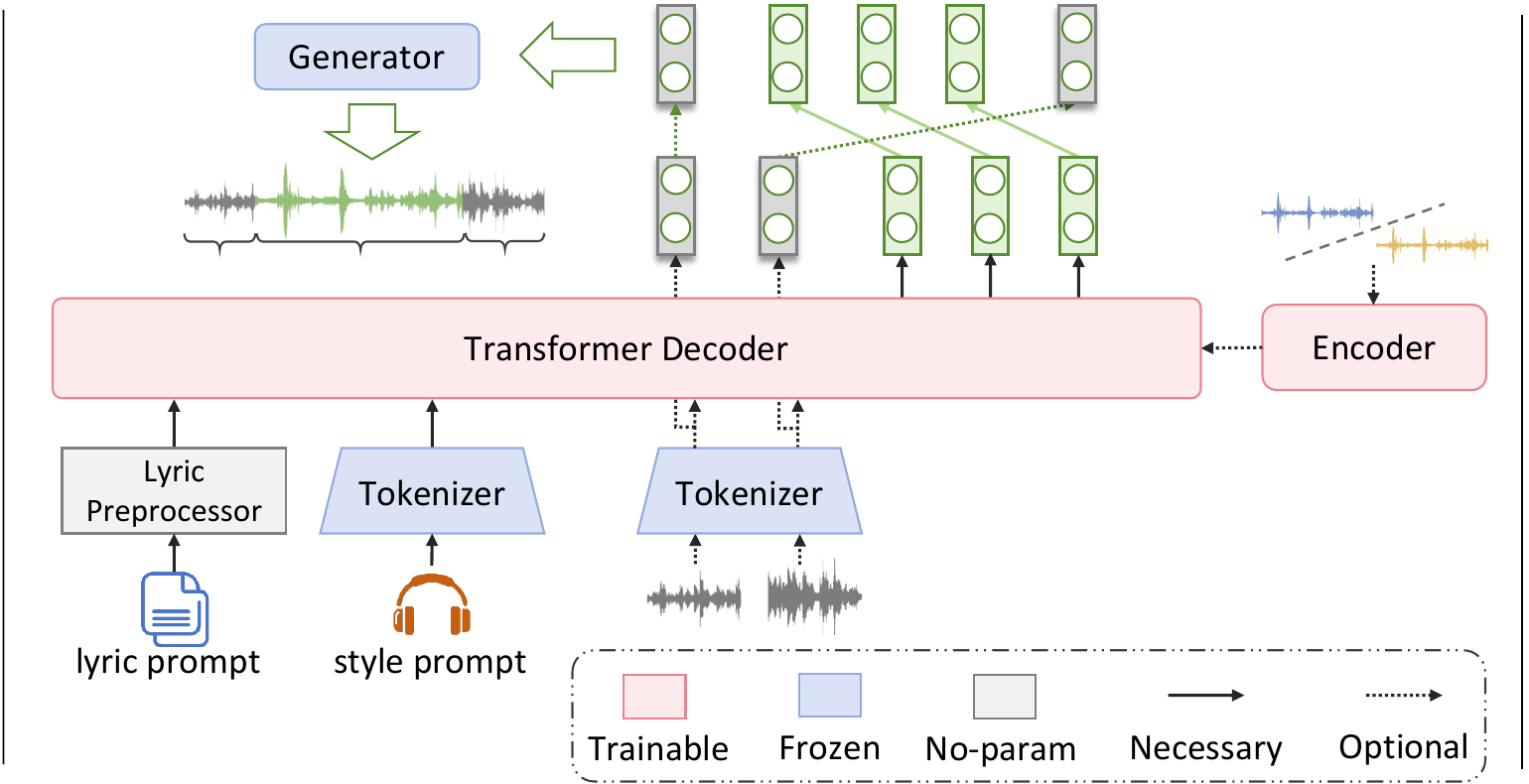}}
    \subfigure[Multi-source encoder for track-wise editing (optional).] {\includegraphics[width=0.23\linewidth]{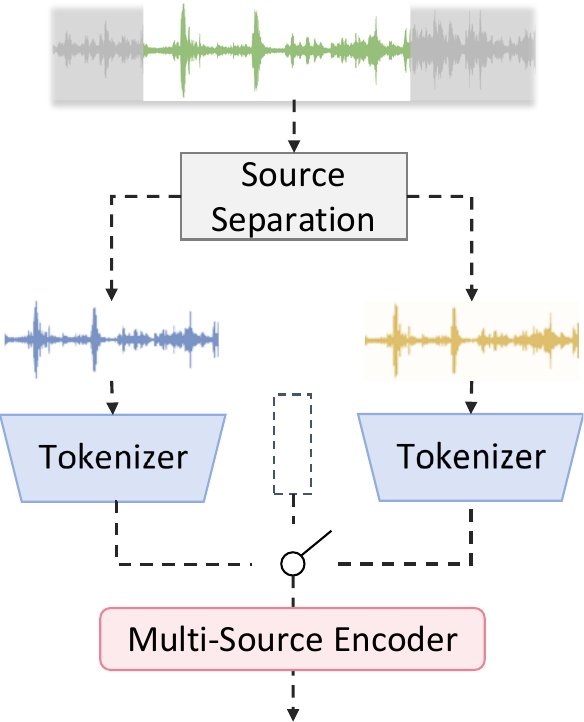}}
    \caption{The architecture of the proposed SongEditor framework. We train the DiT and RVQ jointly first and then the semantic language model. The multi-source encoder is exclusively used for track-wise editing. }
    \label{fig:architecture}
\end{figure*}

\subsection{Zero-Shot Speech Synthesis \& Editing}
Zero-shot Text-to-Speech (TTS) enables generating speech of unseen speakers based on their enrolled recordings. The current mainstream generation methods can be divided into language-modeling approaches and diffusion approaches.
The former~\cite{wang2023neural, du2024vall,borsos2023audiolm} proposes to generate discrete codec tokens~\cite{defossez2022high} autoregressive, while the latter~\cite{shen2023naturalspeech,chen2024f5,yang2024simplespeech,vyas2023audiobox, wang2023audit} leverages diffusion models to generate internal continuous features. 

Moreover, some approaches have extensively explored the task of speech editing. For instance, VoiceCraft~\cite{peng2024voicecraft} proposes a token rearrangement method for the autoregressive generation procedure in neural codec language models. 
Apart from this, diffusion models~\cite{vyas2023audiobox, wang2023audit, du2024unicats} naturally possess the infilling editing ability in a non-autoregressive manner. These approaches have achieved a high quality of speech resynthesis. However, since the speech utterances used in these studies are relatively short and simple, they cannot be directly employed for song editing.
\subsection{Music Generation} 





Recently, music generation has gained substantial attention from the research community. Compared to speech, music typically has a greater degree of variation in pitch, melody, tempo, and timbre. To facilitate user-friendly conditional generation without expert acknowledges, text-to-music has become exceptionally popular. For instance, MusicGen~\cite{copet2024simple} and Mustango~\cite{melechovsky2023mustango} employ the T5~\cite{raffel2020exploring} and Flan-T5~\cite{chung2024scaling} text encoder to process the natural language prompts, and is capable of generating controllable music. Since the paired music and caption data is hard to collect, some approaches~\cite{agostinelli2023musiclm,lam2024efficient, chen2024musicldm,evans2024long} leverage cross-modal alignment models~\cite{huang2022mulan,elizalde2023clap,wu2023large} to project both music and text into the same embedding space instead. The condition vector derived from music can be used to train generative models in a self-supervised manner, while the vector from text is used for practical synthesis purposes.
Some approaches incorporate musical signals, such as chromagrams~\cite{copet2024simple} or beats~\cite{lin2023content} as auxiliary conditions. This additional information assists in making the generation process more controllable.

Several intriguing studies have also focused on music editing. For instance, Music ControlNet~\cite{wu2024music} receives multiple types of musical signals to achieve aime-varying controls. MusicMagus~\cite{zhang2024musicmagus} aims to partially edit specific attributes of generated music by comparing the differences between old and new prompts. Meanwhile, VampNet~\cite{garcia2023vampnet} is designed to complete the masked regions in acoustic token sequences with a non-autoregressive transformer.

\subsection{Singing Voice Synthesis}

Singing Voice Synthesis (SVS) aims to generate vocals based on both lyrics and musical scores. Benefiting from note pitches and durations, the SVS process becomes more controllable and easier to converge.

Conventional SVS systems primarily focus on incorporating phoneme-level melody controls into the generated acoustic representations. For instance, VISinger~\cite{zhang2022visinger,zhang2022visinger2} adopts a methodology similar to VITS~\cite{kim2021conditional}, which learns the alignments between acoustic and textual units. TokSing~\cite{wu2024toksing} discretizes raw audio using multiple self-supervised learning models and further enhances melody information during both the training and inference phases. Recent models~\cite{liu2022diffsinger,hwang2025hiddensinger,he2023rmssinger} have demonstrated promising performance for high-quality singing voice generation. HiddenSinger~\cite{hwang2025hiddensinger}, for instance, utilizes the latent diffusion model~\cite{rombach2022high} to synthesize regularized intermediate codec representations. Meanwhile, RMSSinger~\cite{he2023rmssinger} proposes a DDPM-based method~\cite{ho2020denoising} that synthesizes voice using word-level coarse-grained musical scores, alleviating the heavy demand for manual annotations in real-world scenarios.




\section{Base Model: SongLM} 

Before introducing the editing framework, we would first like to demonstrate the base system, SongLM, for our editing paradigm.
Figure~\ref{fig:architecture} presents the architectural framework of SongLM, which accepts multiple lyric sentences and a 10-second acoustic prompt as inputs. The primary components of this framework include a semantic tokenizer, a language model, and a diffusion-based generator. The semantic tokenizer compresses audio waveforms including the acoustic prompt, context, and separated tracks, into discrete semantic tokens. Subsequently, the language model generates semantic tokens in an autoregressive manner. The final waveform is reconstructed from the output sequence by the diffusion generator.

The training process is divided into two distinct phases. Initially, the tokenizer and diffusion generator are trained concurrently. Subsequently, the language model is trained using tokens generated by the tokenizer. During the second phase, both the tokenizer and the diffusion generator remain in a frozen state.

\paragraph{Lyric preprocessor}

The original data pairs only contain plain text and waveform $\{(X_i, W_i)\}_{i=1}^{L}$, where $L$ is the number of lyrics, $X_i$ is the $i$-th sentence and $W_i$ is the corresponding waveform. It is worth noting that both waveforms and lyrics are consecutive, so they can be concatenated in sequence and $W_i$ may be empty at non-vocal sections. However, relying solely on the text of lyrics is not enough for the complicated song generation task. In order to incorporate the structure information into the lyrics, a structure detector ~\cite{taejun2023allinone} is applied. The category of structure indicators is shown in Appendix A.








\paragraph{Semantic Tokenizer}
Neural codec tokens have been widely used by previous music generation systems~\cite{huang2022mulan}. However, we found that the quantized semantic tokens~\cite{borsos2023audiolm,agostinelli2023musiclm} derived from self-supervised training models exhibit a higher compression rate. 
In this paper, we propose a tokenizer consisting of two distinct branches with different pretrained encoders:  MERT~\cite{li2023mert} and HuBERT~\cite{hsu2021hubert}. The former focuses on accompaniment while the latter is for vocal. Two Residual Vector Quantizers (RVQ)~\cite{zeghidour2021soundstream,kumar2024high} are appended to the end of each branch. And each quantizer has two layers, resulting in $K=4$ tokens in each frame. The semantic token extraction can be formulated as:
\begin{align}
    Y = \mathrm{RVQ}\left(\mathrm{MERT}(W)\right) \oplus \mathrm{RVQ}\left(\mathrm{HuBERT}(W)\right),
\end{align}
where $Y\in \mathbb{Z}^{T\times K}$ is the discrete token sequence and $T$ is the sequence length. $W$ is the input waveform. $\oplus$ denotes the frame-by-frame concatenation.  During the training phase, only the RVQ layers are updated using a commitment loss~\cite{van2017neural}:
\begin{align}
    \mathbb{L}_{rvq} = \sum^K \left( || sg(e_k) - z_k ||^2_2 + || e_k - sg(z_k) ||^2_2 \right),
\end{align}
where $sg(\cdot)$ is the stop-gradient operation, $z_k$ is the input latent vector and $e_k$ is the nearest codebook entry. 

\paragraph{Decoder-only Language Model}
The condition tensors and semantic token sequence are first concatenated together and then fed into a decoder-only transformer. Each transformer layer contains a causal self-attention mechanism with Rotary Position Embeddings (RoPE)~\cite{su2024roformer} and a feed-forward block.

\paragraph{Diffusion Generator}

We utilize the Latent Diffusion Model (LDM)\cite{rombach2022high} as the generator, comprising a diffusion model, a variational autoencoder (VAE)\cite{kingma2013auto}, and a vocoder. We replace the conventional U-Net~\cite{ronneberger2015u} backbone with DiT~\cite{peebles2023scalable}, which conditions on semantic tokens and applies forward and reverse processes on the latent vectors produced by VAE~\cite{van2017neural}. The VAE then converts latent vectors into a Mel spectrogram, which is subsequently transformed into a waveform by the HifiGAN~\cite{kong2020hifi} vocoder.


\section{SongEditor}
\label{sec:4}
%
In this section, we will present SongEditor, a novel framework that integrates editing capabilities into a base model. SongEditor enables two types of song editing: segment-wise, which modifies the vocals and accompaniment within a song segment simultaneously, and track-wise, which allows independent editing of the vocals or accompaniment. Importantly, these approaches can be combined, empowering the flexibility to edit song segments on a single track or holistically.
In the following, we will detail how SongEditor implements these editing functionalities.

\subsection{Long-Content Segment-Wise Editing}

For segment-wise editing, we first discuss leveraging contextual information for the segment being edited. Then, we introduce a rearrangement operation similar to VoiceCraft~\cite{peng2024voicecraft}, applied to the semantic token sequence. To achieve a more natural transition effect, we propose a force-smoothing strategy during training and score-based candidate selection during inference.


\paragraph{Context Selection} 

 The waveform segment to be edited is defined as $W_{[L_A:L_B]} = [W_{L_A}, W_{L_A+1} , \cdots , W_{L_B}] $, where $1\leq L_A \leq L_B \leq L$. Here, $L_A$ and $L_B$ denote the start and end sentences. It is worth noting that $L_A$ and $L_B$ may be equal to 1 or $L$, allowing SongEditor to also perform the continuation or generation task. During the training phase, the preceding and following contexts \( y_{[1:A)} \) and \( y_{(B:T]} \) are directly cut from the token sequence $Y$, where $A$ and $B$ are the start and end indexes of frames. During inference, these contexts are extracted by the semantic tokenizer from the waveforms \( W_{[1:L_A)} \) and \( W_{(L_B:L]} \) respectively, with an overlap of 1 second to avoid edge effects. The overlapped area is subsequently removed after tokenization.
 For the lyrics input, only the edited sentences are retained, represented as $ X = [X_{L_A} , X_{L_A+1} , \cdots , X_{L_B}]$, which we refer to as a lyric context-free strategy. The choice of not including the lyric context is based on two observations: 1) The contextual connection between lyric sentences is not particularly tight. 2) A context-free approach can achieve more precise control over resynthesized vocals, as contextual lyrics can sometimes lead to incorrect vocal generation.


Additionally, we have observed that the context-free strategy is more suitable for practical applications, such as consecutive generation for songs that exceed the maximum length of training samples. By eliminating the need for context lyrics, the language model can automatically generate a much longer song one step at a time~\cite{agostinelli2023musiclm}, without requiring human interception and annotation of the preceding context at each step. This approach enhances both efficiency and usability in practical scenarios.




\paragraph{Rearrangement}
Assuming the original sequence consists of three segments: $Y = [ y_{[1:A)},y_{[A:B]}, y_{(B:T]} ]$, during the training phase, the editing segment is moved to the end of the sequence, and a $\langle$SEP$\rangle$ token is appended at the end of each segment. Thus, the rearranged sequence should be $ Y^{re} = [y_{[1:A)}, y_{<s>}, y_{(B:T]}, y_{<s>}, y_{[A:B]}, y_{<s>}] $, where $ y_i = (y_{i,k})_{k=1}^K $ and  $y_{<s>}$  denotes the embedding of the $\langle$SEP$\rangle$ token.

A delayed pattern~\cite{copet2024simple} is further applied to rearrange the $K$ RVQ tokens. Tokens of the $k$-th quantizer are moved $(k-1)$ timesteps backward and extra $(K-1)$ empty frames are appended to each segment in order to prevent overlap of quantizers. Then tokens at the same timestep are stacked together. Consequently, the final length of the language model input should be $(T + 3K)$. Because the lyrics of contexts have been discarded, only the training loss of the editing segment is calculated as $\mathbb{L}_{ce}=-\log\left(P_{\theta}^{[A,B]}\right)$, where

\begin{align}
    P_{\theta}^{[A,B]} =& P_{\theta}\left(y_{[A:B]}|y_{[1:A)},y_{(B:T]}, X,Y^{sty}\right) \\
     =& \prod_{i=A}^B \prod^K p_{\theta}\left(y_{i,k} | \forall y_{m,n}, X, Y^{sty}\right),
\end{align}
if $m\in\left(0, A\right)\cup\left(B,T\right)$ or $n+m<i+k$.

\paragraph{Force-Smoothing Training}
Due to the presence of melody and rhythm, listeners tend to be more sensitive and strict about interruptions in music compared to speech. However, ensuring smoothness is more challenging for music editing, especially at the endpoint of the editing segment. To tackle this issue, we propose a force-smoothing strategy. During the training phase, the model is enforced to predict additional $\lambda$ frames with a probability of 0.1 for each step even after the editing segment has ended. These frames are directly copied from the beginning of the following context $y_{(B:T]}$, and the $P_{\theta}^{[A,B]}$ for loss calculation is replaced by $P_{\theta}^{[A,B+\lambda]}$. During inference, the model will first greedily determine whether to stop prediction at the current step. As long as the probability of $\langle$eos$\rangle$ is the largest, the model will immediately output $\langle$eos$\rangle$ token. Otherwise, a top-k sampling method will be employed to select the next token randomly from the remaining options. this sampling strategy can effectively alleviate the over-writing issue caused by force-smoothing during training.


\paragraph{Score-Based Candidate Selection}
To achieve better transitions, a score-based candidate selection~\cite{jiang2023mega} is applied during inference. Specifically, we first generate an initial candidate $\hat{y}^1_{[A:B]}$ and resynthesize the last 3 seconds for $(N-1)$ times, resulting in a set of $N$ candidates $\hat{Y}_{[A:B]} = \{ \hat{y}^1_{[A:B]}, \hat{y}^2_{[A:B]}, \cdots, \hat{y}^N_{[A:B]} \}$. Each candidate is used as a new prefix to predict the subsequent $\lambda$ frames. The score of each candidate can be represented as the log-likelihood of $y_{(B:B+\lambda]}$. The candidate with the highest score is ultimately selected, which can be formulated as:
\begin{equation}
    y = {\arg\max}_{\hat{y} \in \hat{Y}_{[A:B]} } P_{\theta}\left(y_{(B:B+\lambda]}|y_{[1:A)},\hat{y}, X,Y^{sty}\right) .
\end{equation}

\subsection{Multi-Source Track-Wise Editing} 
\label{sec:track_edit}
Furthermore, we explore the integration of vocal and accompaniment completion into the model. As depicted in Figure~\ref{fig:architecture}, SongEditor utilizes either the separated vocal or accompaniment track as auxiliary context to complete the other. The provided track is processed through a gated multi-source encoder and incorporated into the language model. Details of this process are introduced below.

\paragraph{Source Separation}
The Band-Split RNN (BS-RNN)~\cite{yu2022high, luo2023music} is adopted as source separation module for track-wise editing. BS-RNN splits the spectrogram into multiple subbands and performs bidirectional LSTM layers across both subbands and frames. Given that music typically has a higher sample rate and a broader frequency range, BS-RNN is exceptionally suitable for music source separation. We leverage BS-RNN to separate vocals from the mixture, leaving the remaining as accompaniments.

During the training phase, a white noise $\epsilon\sim\mathbb{N}(0, \sigma^2)$ is added to the separated vocals in order to prevent the potential leakage of remaining music components after separation~\cite{donahue2023singsong}. In this paper, $\sigma$ is set to 0.01, resulting in a signal-to-noise ratio (SNR) of 40dB.

\paragraph{Multi-Source Encoder}
As shown in Figure~\ref{fig:architecture}, to be compatible with various conditions, both vocal and accompaniment share the same tokenizer and embedding layers. The token embedding of each frame can be represented as:

\begin{equation}
  c_i=\left\{\begin{array}{cc}
       \sum^{K}c_{i,k} + t &t\in \{t^M, t^V\}\\
       t & t = t^{\oslash}.
  \end{array}
  \right.
\end{equation}
where  $t^M$, $t^V$, and $t^{\oslash}$ are special embeddings indicating the type of source, respectively accompaniment, vocal, and none. $c_{i,k}\in \mathbb{R}^D$ is the quantizer embedding and $D$ is the dimension of embedding vectors. If one source is provided, the sequence should be $c=(c_t)_{t=1}^T \in \mathbb{R}^{T\times D}$; otherwise, $c \in \mathbb{R}^{1\times D}$. The sequence is passed through a multi-layer transformer encoder with RoPE and injected into each transformer decoder layer via cross-attention.

\section{Experimental Settings}

\subsection{Datasets}
For the training of SongEditor, a large-scale dataset with approximately 700K songs was used, which sums up to 50K hours. A data cleaning pipeline similar to ~\cite{yu2024autoprep} was adopted to filter out the low-quality samples and correct the timestamp of lyrics in the preprocessing process. BSRNN is initially employed to extract vocals. Pyannote~\cite{Bredin23} and WhisperX~\cite{bain2022whisperx} toolkits are then leveraged for voice activity detection (VAD) and singing voice recognition (SVR) respectively. Lyrics with serious mismatches will be discarded. The time boundary will be corrected based on the VAD results and two adjacent lyric sentences with a short gap will be merged.

For objective evaluation, we assembled a test set of $200$ randomly selected samples. Each sample ranges from $90$ to $120$ seconds in length and contains at least one complete verse and chorus. A $10$-second prompt is extracted from other parts of the song. For the song editing task, we randomly mask a region, which may be located in the middle, beginning, or end of the song. For subjective evaluation, we randomly select $15$ samples with manual correction of annotations, and two recordings are generated based on each sample, which are then assessed by expert musicians.

\begin{table}
\centering
\resizebox{.99\columnwidth}{!}{
\begin{tabular}{l|ccccc}
    \hline
    Method & PER(\%)$\downarrow$ & FAD$\downarrow$ &  Musicality$\uparrow$ & Quality$\uparrow$ \\
    \hline
    GT(restore) & 5.51 & 0.86 &   3.63 & 3.59 \\ \hline
    SongLM & 20.63 & \textbf{1.99} &  2.95 & 3.06 \\
    SongEditor & \textbf{18.33} & 2.24 & \textbf{3.02} & \textbf{3.19} \\
    \hline
\end{tabular}
}
\caption{Evaluation results of baseline and our proposed method for song generation.}
\label{tab:table1}
\end{table}

\begin{table*}[t]
\centering
\resizebox{.99\linewidth}{!}{
\begin{tabular}{l|c|cc|cccc}
    \hline
    \multirow{2}{*}{Method} & \multirow{2}{*}{Context-Free} & \multicolumn{2}{c|}{Objective} & \multicolumn{4}{c}{Subjective (MOS)} \\
    \cline{3-8}
    & & PER(\%)$\downarrow$ & FAD$\downarrow$ &  Musicality$\uparrow$ & Quality$\uparrow$ & Coherence$\uparrow$ & Smoothness$\uparrow$ \\
    \hline
    VoiceCraft~\cite{peng2024voicecraft} & \usym{2715} & 35.23 & 1.34 & 3.03 & 3.17 & 3.48 & \textbf{3.63} \\
    \hline
    SongEditor & \usym{1F5F8} & \textbf{20.82} & \textbf{1.30} & \textbf{3.25} & \textbf{3.39} & \textbf{3.82} & 3.53 \\   
     ~~~ - CS & \usym{1F5F8} & - & - & - & - & - & 3.50 \\   
    ~~~ - CS \& FS & \usym{1F5F8} & 25.68 & 1.37 & 3.13 & 3.23 & 3.54 & 3.43 \\
    \hline
\end{tabular}
}
\caption{Evaluation results of SongEditor for segment-wise editing. “CS” represents the score-based candidate selection. Since it only affects the local smoothness, other metrics are dismissed. “FS” represents force-smoothing.}
\label{tab:table2}
\end{table*}
\subsection{Evaluation Metrics}

To conduct a comprehensive comparison, we evaluated three different models: SongLM, SongEditor, and SongEditor+. SongLM serves as our baseline and can only generate songs from scratch. SongEditor supports segment-wise editing, while SongEditor+ is capable of both segment- and track-wise editing. Details of them are presented in Appendix B.

The proposed models are evaluated in terms of both objective and subjective metrics. For objective evaluation, both Phoneme Error Rate (PER) and Fr\'echet Audio Distance (FAD)~\cite{kilgour2018fr} are employed. To calculate PER, the vocal is first extracted by BSRNN. Then the Whisper-large-v2 is utilized for speech recognition. All punctuation and structure tokens in the lyrics have been removed. The FAD score is computed based on the internal representation from the last layer of MERT-95M, and the fadtk\footnote{\url{https://github.com/microsoft/fadtk}} toolkit is used for calculating statistics and comparison. For editing tasks, only the editing segments are taken into consideration.

For human evaluation, we conduct a mean opinion score (MOS) listening test. Specifically, we employ 30 listeners to rate each audio sample on a scale from 1 to 5 across five aspects: musicality, quality, coherence, smoothness, and intelligibility. Coherence measures the degree of consistency between the editing part and context. Smoothness refers to the naturalness of transitions. These two scores are only considered for the segment-wise editing task.

In our experiments, we observe that tracks restored from the original audio sometimes significantly reduce the performance of the source separation module, resulting in an abnormally high PER. To adjust the quality of vocals more precisely, we also ask the listeners to assess whether the generated vocals sound clear and match the given lyrics (intelligibility) for a track-wise editing task.

\section{Results and Analysis}
\subsection{Song Generation}

Table~\ref{tab:table1} compares the performance of our proposed methods to the baseline for lyric-to-song generation. Additionally, we demonstrate the results of directly restoring audio from ground truth semantic tokens ("restore"), which represents the upper bound of this framework. Incorporating segment-wise editing capability does not degrade the performance of SongLM. Instead, a slight improvement can be observed in terms of subjective MOS scores and PER.

\subsection{Segment-Wise Editing}
Table~\ref{tab:table2} reports the result of SongEditor for segment-wise editing. To demonstrate the advantage of the context-free strategy used in our model,  we adopt VoiceCraft, a context-based approach for speech editing as our baseline. VoiceCraft takes the complete lyrics of both contexts and modified sentences as input. During training, the start and end points are randomly selected. The proposed SongEditor significantly outperforms the baseline in PER. We further find that sentence-level mistakes, such as repeating or missing, are more likely to happen around the editing segments. We believe these mistakes are caused by the failure to align the boundary of context audio and lyrics. In most of the subjective metrics, especially the coherence score, SongEditor also surpasses the baseline,  which proves that the context-free strategy is more suitable for such a long-content editing task. One notable observation is that the smoothness score deteriorates for our proposed model. We attribute this decline to the potential inaccuracies in the annotation of temporal boundaries within the training data.

An ablation study was conducted to verify the effectiveness of force-smoothing training. We retrained SongEditor for the same number of steps without force-smoothing. As shown in Table~\ref{tab:table2}, all metrics declined to varying degrees. This may suggest that force-smoothing training not only enables the model to produce smoother transitions but also helps it learn long-term dependencies in the following context. Additionally, with the score-based candidate selection during inference, SongEditor can achieve a higher smoothness score even without further training.

\begin{figure}
    \centering
    \subfigure[Smoothing.] {\includegraphics[width=0.9\linewidth]{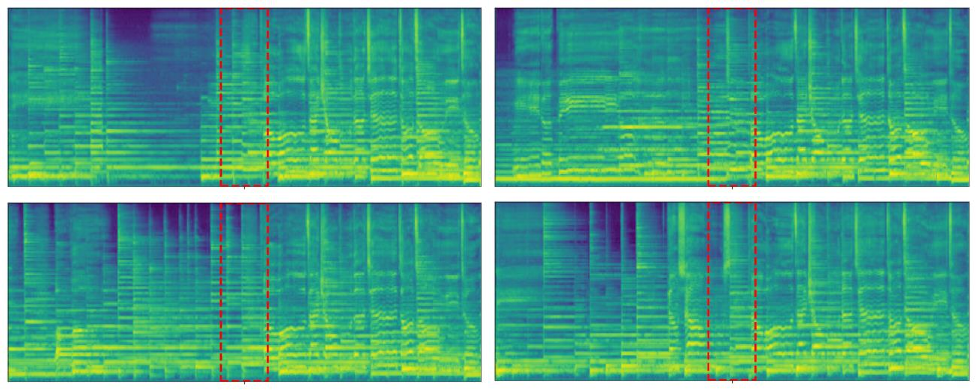}}
    \subfigure[No smoothing.]{\includegraphics[width=0.9\linewidth]{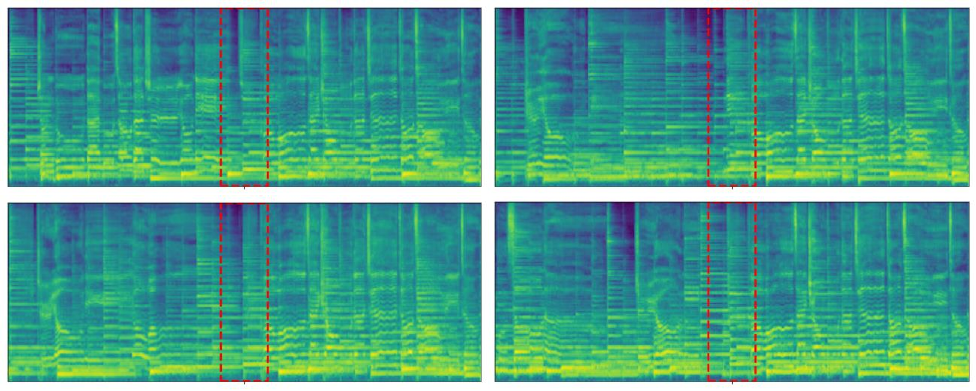}}

    \caption{Mel spectrograms of transitions. The center of the red box is the transition point. The left half is generated while the right half is restored from ground truth.}
    \label{fig:smooth}
\end{figure}

Figure~\ref{fig:smooth} compares the Mel spectrograms of the transition areas. We randomly generated several samples with the same context. Since the autoregressive generation progresses from left to right, the transition from the preceding context to the edited segment is relatively straightforward, so we primarily focus on transitions from generated to subsequent audio. A 10-second audio clip is captured around the end point of each sample. 
As shown in Figure~\ref{fig:smooth},  after applying force-smoothing and candidate selection, the Mel spectrograms of the generated samples appear smoother, with high-energy bands becoming more continuous. In contrast, distinct discontinuities or mismatches can be observed at the transition points without force-smoothing.


\subsection{Track-Wise Editing}

\begin{table*}[t]
\centering
\begin{tabular}{cc|c|cc|ccccc}
    \hline
    \multicolumn{2}{c|}{input}  & \multirow{2}{*}{w/ context}  & \multicolumn{2}{c|}{Objective} & \multicolumn{5}{c}{Subjective (MOS)} \\
    \cline{4-10}
     V & A & & PER(\%)$\downarrow$ & FAD$\downarrow$ & Musicality$\uparrow$ & Quality$\uparrow$ & Coherence$\uparrow$ & Smoothness$\uparrow$ & Intelligibility$\uparrow$ \\

    \hline
     \usym{1F5F8} & \usym{2715} & \usym{2715} & - & 2.71 & \textbf{3.38} & \textbf{3.54} & - & - & \textbf{4.35} \\
     \usym{2715} & \usym{1F5F8} & \usym{2715} & - & \textbf{1.29} & 2.83 & 2.90 & - & - & 3.74 \\   
     \usym{2715} & \usym{2715} & \usym{2715} & 19.30 & 2.61 & 2.68 & 2.92 & - & - & 3.98  \\
    \hline
     \usym{1F5F8} & \usym{2715} & \usym{1F5F8} & - & \textbf{1.13} & \textbf{3.56} & \textbf{3.66} & \textbf{4.19} & \textbf{3.99} & \textbf{4.33} \\
     \usym{2715} & \usym{1F5F8} & \usym{1F5F8} & - & 1.17 & 3.21 & 3.06 & 3.48 & 3.72 & 3.53 \\   
     \usym{2715} & \usym{2715} & \usym{1F5F8} & 23.72 & 1.32 & 2.93 & 3.06 & 3.33 & 3.47 & 3.57 \\
    \hline
\end{tabular}
\caption{Performance of SongEditor+ with different source tracks. “V” is the abbreviation of vocal and “A” is accompaniment. 
The upper part compares the performance for pure generation, while the lower part introduces additional contextual information. }
\label{tab:table3}
\end{table*}

\begin{figure*}
    \centering
    \includegraphics[width=\linewidth]{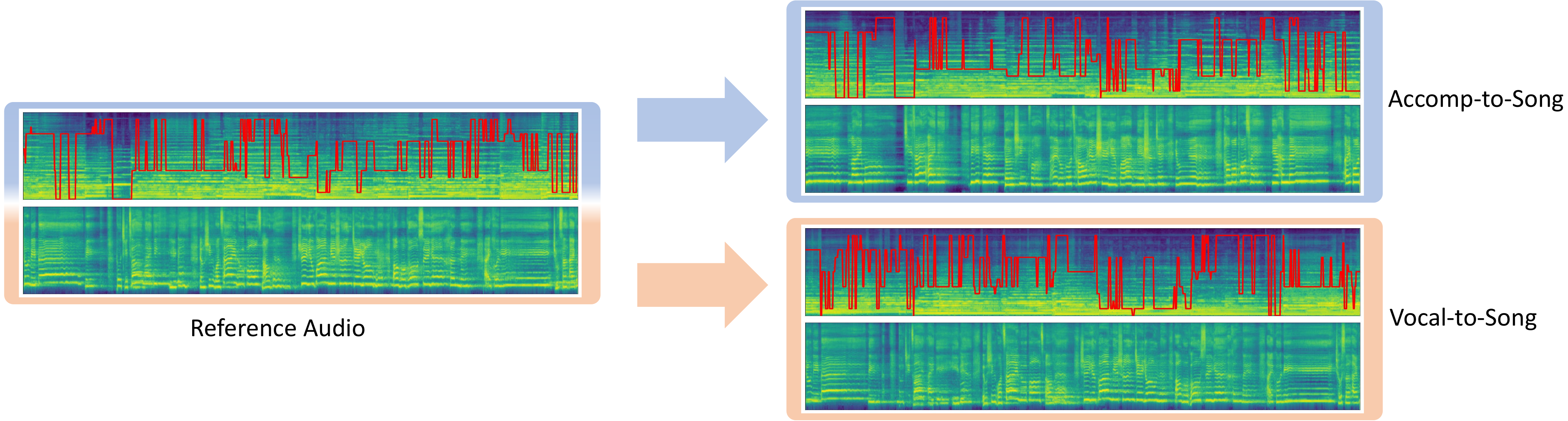}
    \caption{An example of track-wise editing. The Mel spectrogram above corresponds to the separated accompaniment, while the below corresponds to the vocal.  Since the spectrogram of accompaniment is more complex and difficult to identify, its chroma change trend is plotted (red line).}
    \label{fig:demo_track}
\end{figure*}

Moreover, we evaluate SongEditor with cross-attention layers on the track-wise editing task. In this section, we primarily use the Intelligibility MOS score to assess the correctness of the generated vocals, as combining restored and generated tracks can result in abnormal PER outcomes. Our experiments investigate the influence of different source tracks by testing various configurations: vocals only (vocal-to-song), accompaniment only (accomp-to-song), and neither of them. Track-wise information is integrated into both the generation and segment-wise editing tasks.

As shown in Table~\ref{tab:table3}, both vocals and accompaniments boost performance in terms of musicality and FAD scores. The vocal track plays a dominant role in track-wise editing, which is expected since vocals often serve as the backbone of a song, directly conveying crucial elements such as rhythm and emotion. Additionally, listeners naturally focus more on vocals.
Conversely, generating vocals from accompaniments presents greater challenges, as the model must maintain precise rhythm synchronization with accompaniments, resulting in a slight decrease in the intelligibility score as a trade-off. Notably, the FAD score for vocal-to-song generation from scratch is considerably high (2.71) but drops significantly when contextual information is provided (1.13). This suggests that the model can learn more detailed accompaniment features when contexts are available.

Figure~\ref{fig:demo_track} illustrates the result of music separation for both reference and generated audio. Our goal is to study the consistency of the regenerated audio with the original source input. Specifically, we first separate the reference audio into vocals and accompaniment, then use SongEditor+ to independently complete each track. As shown in the figure, the generated vocal spectrogram closely resembles the input in the vocal-to-song task. Conversely, for accomp-to-song, while the restored chromagram generally remains consistent, some differences can be observed in both the chromagram and spectrogram. We believe that vocals are relatively independent and easier to compress, while accompaniments often rely on vocals and contain more detailed information.

\subsection{Consecutive and Multi-Singer Story Mode}

As highlighted in Section 4, the context-free strategy demonstrates significant advantages in practical applications. In this section, we explore generating full songs longer than two minutes through an iterative round-by-round process, referred to as Story Mode, following \citet{agostinelli2023musiclm}. The complete lyrics are pre-divided into several sections. During each round, only the lyrics of the current section are provided, while a fixed-length token sequence is automatically extracted from the previous output as the prefix. No manual intervention is required throughout the process.

Specifically, we adopt two task settings: full song generation and multi-singer story mode. The former aims to generate a full song from intro to outro, without changing the style prompt. This task focuses on the coherence across rounds, and we advance with a stride of 60 seconds. For the latter, we use two different style prompts, typically one for male and one for female. Verse and chorus sections are generated with alternate prompts. In order to prevent the emergence of vocal in the prefix, a short instrumental segment is generated at the end of each round and the stride length is set to 5 seconds.
Experiments demonstrate that our model can generate full songs with smooth transitions and coherent content.  By slightly modifying the task settings, it can also achieve multi-singer generation with variational style prompts. Audio samples are presented in the demo page.

\section{Limitations and Ethic Discussion}

\noindent \textbf{Limitations:} SongEditor currently regenerates specified sentences without providing additional control over the editing segment. Furthermore, there is no explicit decoupling of tracks at the semantic level, which could enhance interpretability and improve the distinction.

\noindent \textbf{Ethics:} SongEditor is an innovative tool that helps either a professional musician or a passionate enthusiast to create their own songs. However, we also fully acknowledge the potential ethical risks. We ensure that our training data does not infringe on any copyright, and only public-domain melodies are used for inference. 

\section{Acknowledgments}
This work was supported by Shenzhen Science and Technology Program (Shenzhen Key Laboratory Grant No. ZDSYS20230626091302006) and Shenzhen Science and Technology Research Fund (Fundamental Research Key Project Grant No. JCYJ20220818103001002).

\bibliography{aaai25}

\clearpage
\section{Appendix A: Lyric Preprocessor}
\label{apx:a}
\begin{figure}[h]
  \centering
  \includegraphics[width=\linewidth]{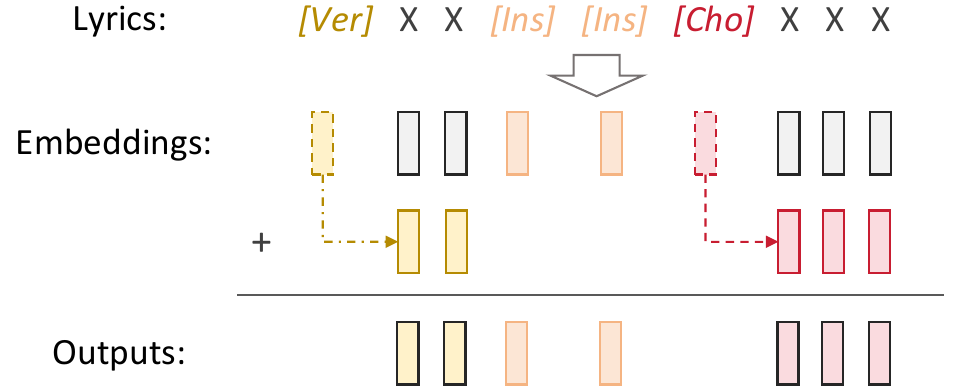}
  \caption{Text token representation for the structure-guided lyrics.}
  \label{fig:lyric}
\end{figure}

Specifically, within this work's context, the song structure is simplified and represented as 6 categories: $[verse]$, $[chorus]$, $[bridge]$, $[intro]$, $[outro]$, $[inst]$, among which 
 $[verse]$, $[chorus]$ and $[bridge]$ indicate the attributes of lyrics (\textit{lyric-related}), while the other tokens describe the non-vocal sections of a song (\textit{accompaniment-only}). To enhance the controllability of structure tokens, these two different types are integrated into the text sequence in different ways, as illustrated in Figure~\ref{fig:lyric}.

 For lyric-related tokens, an addition operation is applied between the type embedding and the corresponding text similar to \citet{devlin2018bert}. For accompaniment-only tokens, each entry is repeated according to the duration of the corresponding non-vocal section and directly inserted into the text sequence.  In this way, a potential alignment between lyrics and semantic tokens is ensured, as each entry in the text sequence corresponds to a duration within the generated song.

\label{secL lrc_proc}

\section{Appendix B:  Configuration Details}

For the semantic tokenizer, we take the output of the 13th layer of MERT-330M and the last layer of HuBERT as semantic representations. And each RVQ has a codebook size of 10,000. The frame rate of semantic tokens is 25.

To conduct a comprehensive comparison, we trained three different models: SongLM, SongEditor, and SongEditor+. SongLM serves as our baseline and can only generate songs from scratch. SongEditor supports segment-wise editing, while SongEditor+ is capable of both segment- and track-wise editing tasks. All models are based on the Llama architecture~\cite{touvron2023llama} with 1024 hidden dimensions and 16 attention heads in each layer with a frame rate of 25. SongLM and SongEditor consist of 16 Llama decoder layers with causal masking. For SongEditor+, an additional cross-attention is inserted into each layer, and the multi-source encoder is a two-layer transformer encoder. Due to computing resource limitations, we reduced the number of layers to 12. Each model contains approximately 800M parameters. The top-k sampling with $k=250$ and the classifier-free guidance~\cite{ho2022classifier} with a coefficient of 1.5 is employed during inference.

We trained all models using the AdamW optimizer~\cite{loshchilov2017decoupled} with a learning rate of 1e-4 for approximately 200K steps. 16 NVIDIA A100 GPUs were used in the training process and the actual batch size is 128. The DeepSpeed ZeRO strategy~\cite{rajbhandari2020zero} and bf16 precision were adopted to accelerate the training and reduce memory consumption. Each condition was dropped with a probability of 0.2 during the training phase. For SongEditor+, each source (vocal, accompaniment, or empty embedding) was randomly selected with equal probability in each batch.

\end{document}